# Specific Heat of $Sr_4Ru_3O_{10}$

X.N. Lin, V.A. Bondarenko[a,], G. Cao, and J.W. Brill[*]

Department of Physics and Astronomy, University of Kentucky, Lexington, KY

40506-0055, USA

ABSTRACT:

We have measured the specific heat of single crystals of the triple-layer Ruddlesden-Popper material, $Sr_4Ru_3O_{10}$, grown both in an image furnace and by flux-growth. The flux grown sample has a sharp mean-field-like anomaly at the onset of magnetic order, $T_C$ = 102 K, but a much broader anomaly, indicative of residual heterogeneity, is observed for the image furnace sample. Even for the flux grown sample, however, the anomaly is at least an order of magnitude smaller than one would expect for complete ordering of the spins. Neither sample exhibits an anomaly at $T_M$ ~ 50 K, where magnetic measurements suggest that basal plane antiferromagnetism sets in. Anomalous behavior (e.g. consistent with a term in the specific heat $\propto T^{3/2}$ as would be observed for a three-dimensional ferromagnet with weak exchange) is observed at low temperatures for both samples, indicative of the unusual magnetic order in this material.

PACS: 75.40.Cx, 71.27.+a

Keywords: A. magnetically ordered materials, D. heat capacity, D. phase transition

*Corresponding author:   E-mail address: jwbrill@uky.edu

Tel: 1-859-257-4670     FAX: 1-859-323-2846

The Ruddlesden-Popper ruthenates, $(Ca,Sr)_{n+1}Ru_nO_{3n+1}$, with n neighboring layers of corner sharing (distorted) $RuO_6$ octahedra separated by alkaline earth layers, have generated considerable interest because of the variety of magnetic and superconducting states they exhibit [1,2], reflecting competition between spin, orbital, and charge ordering in these materials. For example, triple layered $Sr_4Ru_3O_{10}$ has magnetic and resistive anomalies at $T_C \approx 102$ K and $T_M \approx 50$ K [3]. $T_C$ marks the onset of apparent ferromagnetic order, with the easy axis normal to the $RuO_6$ layers, but with saturated moments markedly less than the expected value of two Bohr magnetons per (S=1) $Ru^{4+}$ ion. At $T_M$, the moments normal to the layers increase somewhat; however, the in-plane moments decrease toward zero, exhibiting a metamagnetic transition at a field of a few Tesla and suggesting that at zero field there is antiferromagnetic-like order for the in-plane spins. However, the magnetic properties, reflecting the complex competition between (in-plane) antiferromagnetic and (interplane) ferromagnetic interactions, are unique in that there appears to be an "easy-plane", rather than an easy-axis, for the antiferromagnetic order. The interlayer resistivity drops by an order of magnitude below $T_M$, presumably due to a reduction in spin scattering, but, reflecting the layered structure, stays more than an order of magnitude greater than the basal plane resistivity [3]. Surprisingly, only in the interplayer resistvity are Shubnikov-de Haas oscillations discerned [3].

In this paper we report on the zero-field specific heat of single crystals of $Sr_4Ru_3O_{10}$. While a small, mean-field like anomaly ($\Delta C \sim 0.4$ R, where R = 8.314 J $mol^{-1}$ $K^{-1}$ is the gas constant) is seen at $T_C$, no anomalies ($\Delta C < 0.05R$) are observed at $T_M$. In addition, a contribution to the heat capacity $\propto T^{3/2}$ is observed at low temperatures, signaling the ambiguity of the ground state mentioned above [3].

Measurements were made both on crystals grown in an image furnace and crystals from reference [3], grown using a flux technique. The structures and stoichiometries of the samples were verified by results of powdered crystal x-ray diffraction and electron dispersive x-ray (EDX). The magnetic and transport properties, measured with a Quantum Design MPMS-LX system (SQUID) with an added four-lead resistivity function, were also used to characterize the samples. While some of the relevant magnetic and transport properties are mentioned below, they will be discussed more extensively in a later publication.

Specific heats were measured by ac-calorimetry using chopped light as an oscillating heat source [4,5]. For temperatures above ~ 15 K, the crystals were attached to fine thermocouple thermometers, while for temperatures between ~ 3 K and 20 K the crystals were attached to Cernox bolometers [5]. Measurements were made on a 0.4 mg flux grown crystal at high temperatures, two flux grown crystals totaling 1.3 mg at low temperatures, and a 1.8 mg image furnace grown crystal at all temperatures. Chopping frequencies were chosen to be between the measured internal and external thermal relaxation rates for each temperature interval, in which case the magnitude of the oscillating temperature is inversely proportional to the total heat capacity of the sample and addenda (thermometer, glues, connecting wires) [4]. Typical frequencies used were ~ 2 Hz at high temperatures and ~ 20 Hz at low temperatures.

Because the absorbed power is not known, our technique only yields relative values of the heat capacity. To normalize these results, the heat capacities of polycrystalline samples of both flux grown and image furnace grown samples were measured using differential scanning calorimetry (DSC) [6]. The values measured with DSC were consistent (within ± 3%), and the ac-calorimetry results were normalized to the DSC values at 200 K, while approximately correcting for the heat capacities of the addenda, which were less than 5% of the sample heat capacities at all temperatures.

Figure 1 shows the specific heat of both the image furnace (IF) and flux grown (FG) samples over the whole temperature range. The close similarity of the temperature dependences of the specific heats suggests that the IF sample has a negligible amount of other strontium ruthenate phases present.

Nonetheless, the results at the Curie point, shown in an enlarged scale in Figure 2, suggest that the image furnace grown sample is more heterogeneous than the flux grown sample. For example, the IF sample may have (non-uniform) strains caused by surface tension as it crystallizes in the floating zones of the furnace and/or small amounts of other phases (e.g. ferromagnetic cubic $SrRuO_3$ [7]). Whereas the flux grown sample exhibits a sharp mean-field-like step in the specific heat, $\Delta C \sim 0.4R$, at $T_C = 102$ K, the anomaly is broadened by ~ 5 K for the image furnace grown sample. Similarly, while the basal plane susceptibility of the FG sample has a sharp cusp at $T_C$ (see Figure 2a of reference [3]), the IF sample has a less pronounced slope change, as shown in the inset to Figure 2. Other differences obvious from a comparison of the Figure 2 inset and Reference [3] are that the susceptibility of the IF sample is much less anisotropic than the FG sample and has a larger magnetization above $T_C$. These differences cannot be simply attributed to impurities in the IF sample, as its (basal plane) residual resistivity ratio (RRR = 14) is close to that of the FG sample (RRR = 21). Instead, the presence of strains and/or small inclusions of other phases in the IF sample may affect the magnetization of the host, allowing some spins to order above $T_C$, but with the resulting heterogeneity broadening the transition. (However, it should be noted that, in contrast to the specific heat results, the widths of the anomalies at Tc, as measured by the large increases in c-axis magnetic moment, are ~ 5K for both the IF and FG samples.) At lower temperatures, the "1D-ferromagnetic, 2D-antiferromagnetic" order intrinsic to $Sr_4Ru_3O_{10}$ again apparently prevails in the IF sample, as shown in the Figure 2 inset.

Even for the FG sample, the specific heat anomaly is much smaller than expected; for example, per ruthenium ion, the anomaly is about an order of magnitude smaller than that observed at $T_C$ in the itinerant ferromagnetic $SrRuO_3$ [8]. One expects an entropy change of $k_B\ln(3)$ for each ordering spin, where $k_B$ is Boltzman's constant; if all three spins/formula unit order, the expected entropy change is therefore $\Delta S = 3.3$ R. Estimating the entropy change from the temperature dependence of the specific heat at such high temperatures is difficult because one doesn't know the correct baseline behavior. An approximate lower limit on $\Delta S$ can be obtained by fitting the specific heat away from the transition to a smooth curve, as shown in Figure 2, and measuring the area under the anomaly; in this case, we obtain $\Delta S > 0.02R$. On the other hand, an upper limit can be estimated from $\Delta S \sim \Delta C$ (~ 0.4R), as expected for a mean-field anomaly. However, even in this case, the entropy change is an order of magnitude smaller than expected for complete spin ordering, suggesting that either the spin ordering is not spatially uniform or that only a small component of the spins order. It is noted that the magnetic entropy removal at $T_C$ is generally small for weakly ferromagnetic metals, chiefly due to spin fluctuations [8,9].

Figure 3 shows the specific heats near $T_M \sim 50$ K. No structure in the specific heat ($\Delta C < 0.05$ R) is observed for either sample, strongly suggesting that structure observed in the magnetic and transport properties at this temperature (peaks in interplane resistivity and basal plane moment, and increase in interplane moment [3]) reflect gradual "crossover" behavior, as the spin order changes, rather than a thermodynamic phase transition, at least in zero field. (Note that the magnetic anomalies at $T_M$ for the FG sample [3] are somewhat sharper than those of the IF sample shown in the Figure 2 inset.) In particular, given the close competition between ferromagnetism and antiferromagnetism, these changes in transport and magnetic properties at $T_M$ may merely signify a change in the relative strength of ferromagnetism and antiferromagnetism.

Figure 4 shows the low temperature specific heats; quantitative differences in the two samples at these temperatures are magnified by the fact that they were normalized near room temperature, so that small errors at intermediate temperatures (e.g. due to marginally correct chopping frequencies, incorrect addendum subtractions) get amplified here. Nonetheless, in C/T vs $T^2$ plots, both samples exhibit similar negative curvature for $T < 9$ K. Such curvature implies that, in addition to the usual phonon ($C_{ph} = \beta T^3$) and electronic ($C_e = \gamma T$) contributions to the specific heat, there must be a term $C_S \sim \eta T^p$ with $1<p<3$, presumably due to spin excitations. Furthermore, for the fits to give (per atom) Debye temperatures below ~500 K, we must have $p < 2$. (If one estimates $\beta$ from the linear portions of the curves only, i.e. from the data for $T > 9$ K, one obtains an average Debye temperature for the two samples of $\Theta = (367 \pm 15)$ K, close to the values for cubic (i.e. $n = \infty$) $SrRuO_3$ and double layer $Sr_3Ru_2O_7$ [8,10,11]. Therefore, including a term with $p<3$ decreases the value of $\beta$ and increases $\Theta$.) These constraints on p suggest magnetic order in apparent contradiction to the measured magnetic properties, again indicating the complexity of the magnetic order.

For example, Figure 4 shows fits with p = 3/2, the value appropriate for three-dimensional ferromagnetic order [12]; while the magnetic structure is obviously more complicated than this, it is still interesting to consider the parameters of the fit. The values of $\beta$ from the fits are 0.037 mJ mol$^{-1}$K$^{-4}$ and 0.052 mJ mol$^{-1}$ K$^{-4}$ for the FG and IF samples, respectively, giving an average value of the Debye temperature of 423 ($\pm$ 24) K. The values of $\gamma$ are 69 mJ mol$^{-1}$ K$^{-2}$ and 46 mJ mol$^{-1}$ K$^{-2}$, for the FG and IF samples respectively, considerably smaller than the value (109 mJ mol$^{-1}$ K$^{-2}$) we assumed in Reference [3] when the negative curvature was not known. The Wilson ratio obtained with our average fitted value of $\gamma$ and the interplane susceptibility [3] is ~ 4.5, reinforcing the importance of electronic correlations in this material [3]. However, we note that our *per Ruthenium* value $\gamma/3 = (19 \pm 4)$ mJ mol(Ru)$^{-1}$ K$^{-2}$ is smaller than the values generally observed in other Ruddlesden-Popper ruthenates (typically 30-100 mJ mol(Ru)$^{-1}$ K$^{-2}$ [7,8,10,11]. Finally, the values of the magnon coefficient in the fits for the two samples are very close: $\eta_{FG}$ = 44 mJ mol$^{-1}$ K$^{-5/2}$ and $\eta_{IF}$ = 43 mJ mol$^{-1}$ K$^{-5/2}$. For a three-dimensional ferromagnet, with three spins/formula unit, $\eta/R$ = 3 (0.040) (SJ$_{eff}$)$^{-3/2}$, where J$_{eff}$ is the effective exchange interaction averaged over nearest neighbors [12]. Hence, taking S = 1, our fit gives SJ$_{eff}$ = 8 K. This small value suggests that if there is residual three-dimensional ferromagnetic order, it affects only a small component of the spins. (Note that a T$^{3/2}$ term with such a small value of J$_{eff}$ cannot be associated with inclusions of ferromagnetic SrRuO$_3$, for which such a contribution was not observed [7,8].)

On the other hand, the magnetic data suggests that there is unusual ferromagnetic interplane and antiferromagnetic intraplane order.  For *conventional* antiferromagnets, the presence of an easy axis for spin alignment results in an antiferromagnetic magnon excitation gap $\varepsilon_A$, so that, for $k_BT \ll \varepsilon_A$, the magnon specific heat would be activated: $C_S \propto T^{1/2}\exp(-\varepsilon_A/k_BT)$, where the prefactor is the contribution of quasi-one-dimensional ferromagnetic magnons [12].  On the other hand, if $k_BT \gg \varepsilon_A$, the combination of quasi-one-dimensional ferromagnetic order and two-dimensional antiferromagnetism would give $C_S \propto T^{5/2}$ [12].  Hence, our results may be consistent with planar antiferromagnetism coupled to 1D ferromagnetic order if the antiferromagnetic anisotropy gap were comparable to temperature, e.g. $\varepsilon_A / k_B \sim 5$ K, so that our apparent $p\sim3/2$ behavior is due to a cross-over between activated and $p=5/2$ behavior.  However, since the anisotropy for the present case is unconventional, in that a large field is required for a metamagnetic transition but there is apparent isotropy within the plane, it is not clear how to interpret this anisotropy gap.

Finally, we mention that the low temperature specific heat can also be fit (as shown by the dashed curves in Figure 4) with a spin term $C_S = \eta T \ln(T/T_0)$, as for ferromagnetic fluctuations near at quantum critical point [13].  The average Debye temperature for these two fits is $\Theta = (395 \pm 20)$ K.  However, since the material exhibits ferromagnetic interplane order, it is again not clear how to interpret such a dependence.

In summary, we've measured the specific heat of crystals of triple-layered $Sr_4Ru_3O_{10}$ prepared both by flux growth and in an image furnace.  While the overall temperature dependence of the samples is very similar, the flux grown sample has a much sharper anomaly at the Curie point $T_C = 102$ K; however even this anomaly is an order of magnitude smaller than expected for complete spin ordering. Neither sample exhibits a thermal anomaly at $T_M \approx 50$ K, suggesting that the magnetic and resistive changes at this temperature are caused by a gradual crossover in spin order rather than a thermodynamic phase transition. Finally, the specific heat at low temperature (3 K < T < 9 K) has an apparent $\sim T^{3/2}$ contribution. While it is not clear how to interpret this dependence, it is presumably symptomatic of the unusual magnetic order in this material. The complex behavior presented points to a possible exotic state where there exists a delicate balance between order and fluctuations.

This research was supported by the U.S. National Science Foundation, grants # DMR-0100572 and DMR-0240813.

**Figure Captions**

Figure 1. Temperature dependence of the specific heats, over the whole temperature range, for the flux grown (FG) and image furnace (IF) samples  The specific heats are normalized to the gas constant, R= 8.31 J mol$^{-1}$ K$^{-1}$, and the data for the IF sample are shifted up by 5 for clarity.

Figure 2. Temperature dependence of C/R for both samples near $T_C$ = 102 K; the data for the IF sample are shifted up by 1 for clarity.  The curve through the FG data is a quadratic function of temperature used to estimate the entropy change, as discussed in the text.   Inset: Temperature dependence of the basal plane and c-axis magnetic moments of the IF sample, measured in a field of 100 Oe.

Figure 3.  Temperature dependence of C/R heat of both samples, normalized to the gas constant, near $T_M$ = 50 K.  The data for the IF sample have been shifted up by 1 for clarity.

Figure 4.  C/RT vs. T$^2$ for both samples; the data for the IF sample have been shifted up by 0.01 K$^{-1}$ for clarity.  The solid curves show fits to $C = \gamma T + \beta T^3 + \eta T^{3/2}$ and the dashed curves show fits to  $C = \beta T^3 + \eta T \ln(T/T_0)$, discussed in the text.

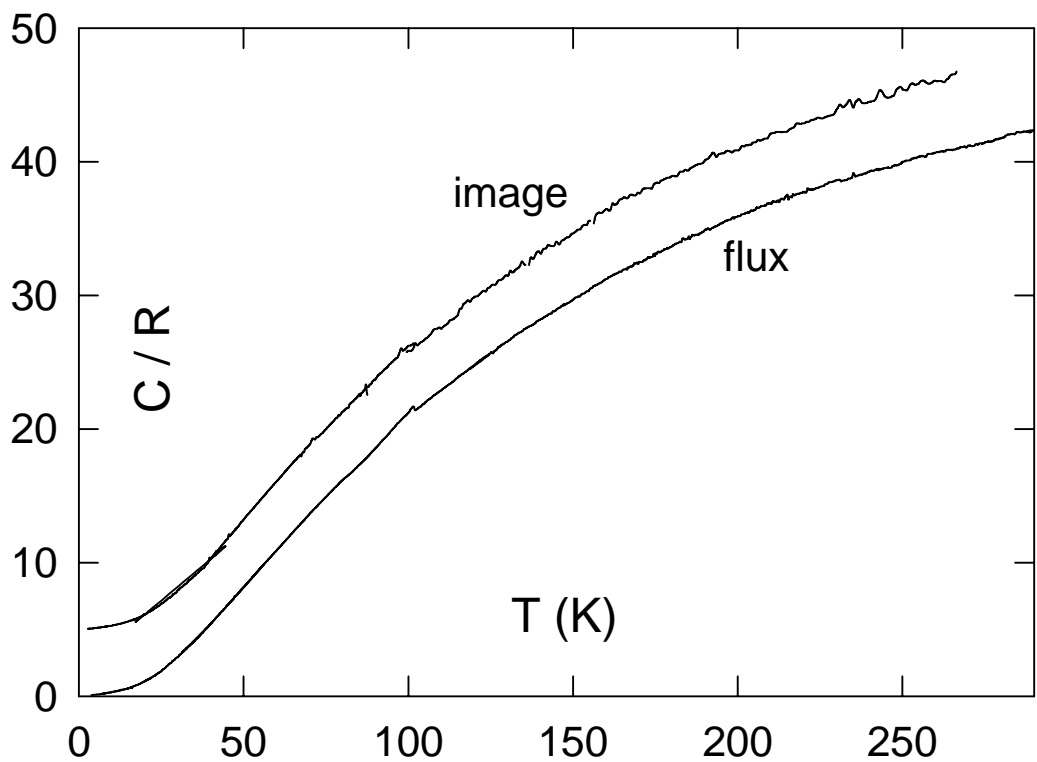

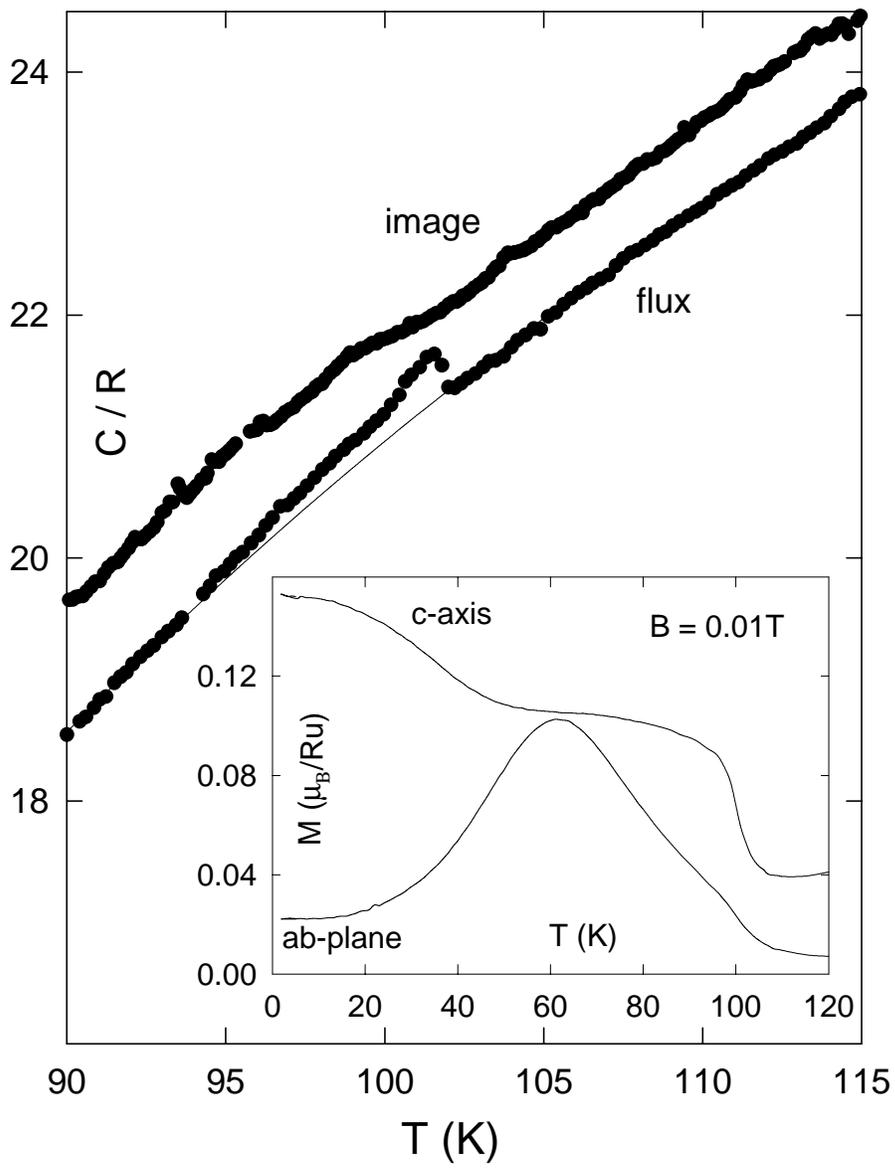

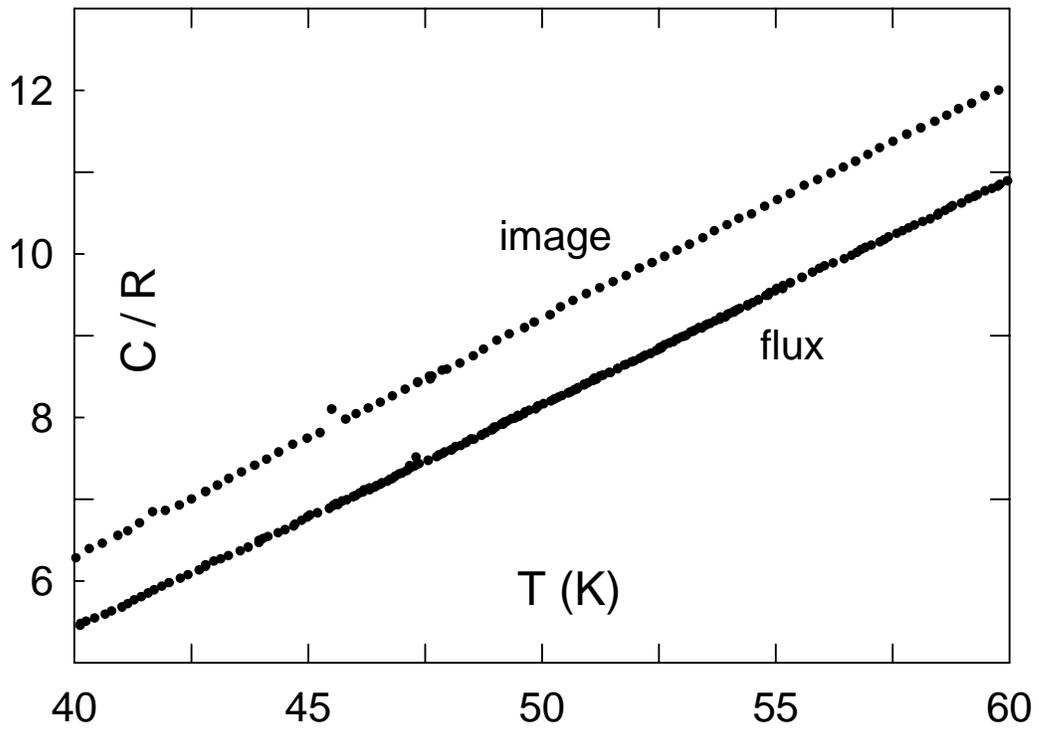

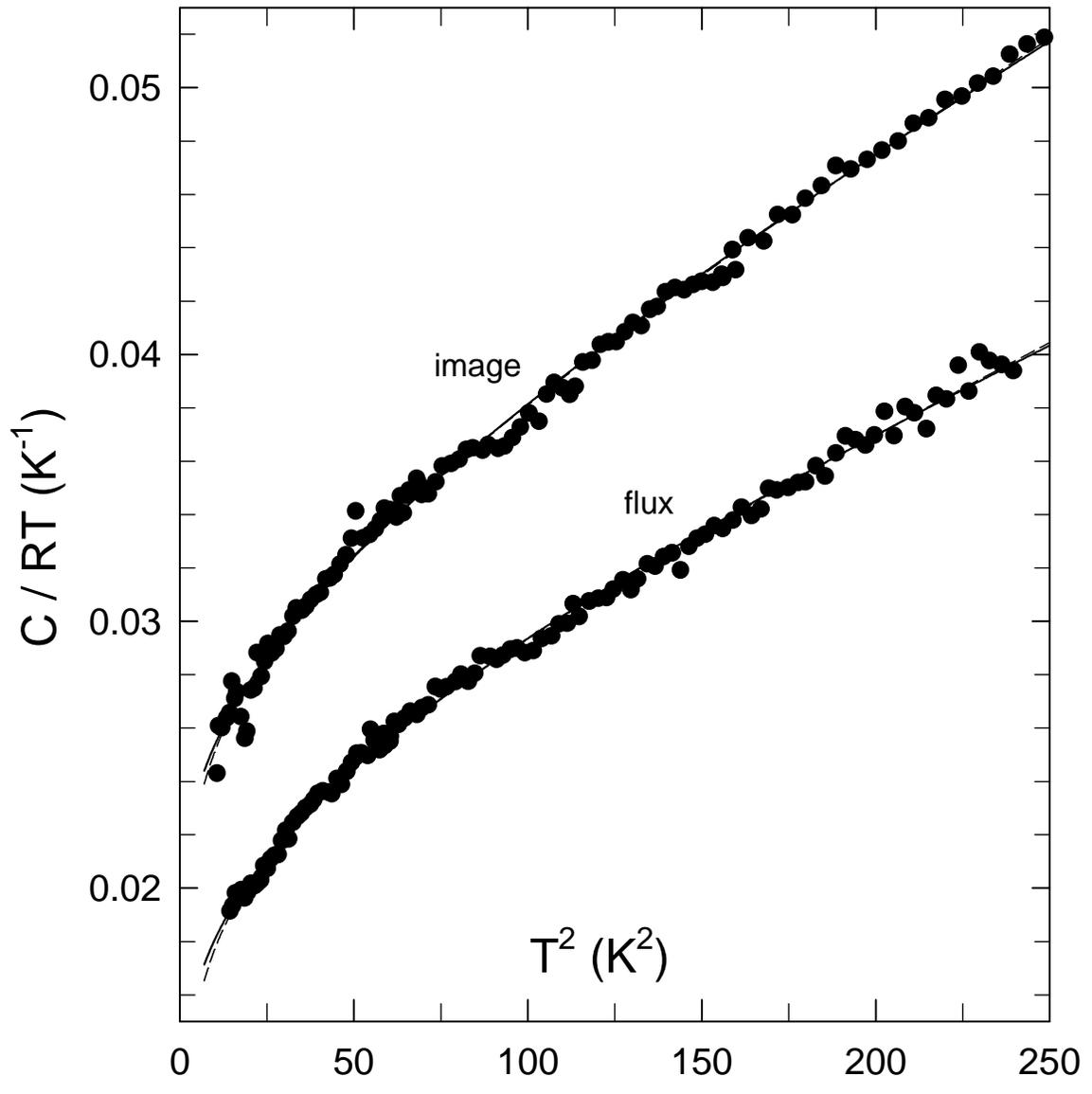